\documentclass[12pt]{article}
\usepackage{amsmath}
\usepackage{graphicx}
\usepackage{natbib}
\usepackage{url} % not crucial - just used below for the URL 
\usepackage{soul,color} % use \hl to highlight text
\usepackage[noend]{algpseudocode}
\usepackage{algorithm}
\RequirePackage[colorlinks,citecolor=blue,urlcolor=blue]{hyperref}%% uncomment this for coloring 

\newcommand{\blind}{0}

\begin{document}

\def\spacingset#1{\renewcommand{\baselinestretch}%
{#1}\small\normalsize} \spacingset{1}

%%%%%%%%%%%%%%%%%%%%%%%%%%%%%%%%%%%%%%%%%%%%%%%%%%%%%%%%%%%%%%%%%%%%%%%%%%%%%%

\if0\blind
{
  \title{\bf Minor Issues Escalated to Critical Levels in Large Samples: A Permutation-Based Fix}
  \author{
    Xuekui Zhang %\orcid{0000-0003-4728-2343} 
    \thanks{This research was funded by NSERC DG: \# RGPIN-2022-03050 (XZ), \# RGPIN-2021-03530 (LX) and the Canada Research Chair \#CRC-2021-00232 (XZ), Michael Smith Health Research BC Scholar: \# SCH-2022-2553(XZ). This research was enabled in part by computational resource support provided by Westgrid (https://www.westgrid.ca) and the Digital Research Alliance of Canada (https://alliancecan.ca).}\hspace{.2cm}\\
    Department of Mathematics and Statistics, University of Victoria,  Canada\\
    and \\
    Li Xing \\ %\orcid{0000-0002-4186-7909) 
    Department of Mathematics and Statistics, University of Saskatchewan, Canada\\
    and \\
    Jing Zhang  \\
    Department of Mathematics and Statistics, University of Victoria,  Canada\\
    and \\
    Soojeong Kim  \\
    Department of Mathematics and Statistics, University of Victoria,  Canada
    }
  \maketitle
} \fi

\if1\blind
{
  \bigskip
  \bigskip
  \bigskip
  \begin{center}
    {\LARGE\bf Title}
\end{center}
  \medskip
} \fi

\bigskip
\begin{abstract}
In the big data era, the need to reevaluate traditional statistical methods is paramount due to the challenges posed by vast datasets. While larger samples theoretically enhance accuracy and hypothesis testing power without increasing false positives, practical concerns about inflated Type-I errors persist. The prevalent belief is that larger samples can uncover subtle effects, necessitating dual consideration of p-value and effect size. Yet, the reliability of p-values from large samples remains debated.\\
This paper warns that larger samples can exacerbate minor issues into significant errors, leading to false conclusions.  Through our simulation study, we demonstrate how growing sample sizes amplify issues arising from two commonly encountered violations of model assumptions in real-world data and lead to incorrect decisions. This underscores the need for vigilant analytical approaches in the era of big data. In response, we introduce a permutation-based test to counterbalance the effects of sample size and assumption discrepancies by neutralizing them between actual and permuted data. We demonstrate that this approach effectively stabilizes nominal Type I error rates across various sample sizes, thereby ensuring robust statistical inferences even amidst breached conventional assumptions in big data. \\
For reproducibility, our R code will be publicly available at: \url{https://github.com/ubcxzhang/bigDataIssue} before April 2024.
\end{abstract}

\noindent%
\vfill

\newpage
\spacingset{1.75} % DON'T change the spacing!
\section{Background}
\label{sec:intro}
In the era of big data, analyzing datasets with vast volumes of entries prompts a reevaluation of the applicability of traditional statistical methods to such extensive data. As a basic fact stated in statistics textbooks, larger sample sizes are theoretically favoured for increased accuracy, reduced variability, and enhanced hypothesis testing power without intensifying Type I error rates. However, in practical scenarios, especially in medical studies, there's a concern that large sample sizes could inflate Type I errors, a point not fully explored regarding underlying causes \citep{10.1111/opo.12618, 10.1590/2176-9451.19.4.027-029.ebo}. A widely accepted view is that larger samples can reveal even minor effects, making it crucial to consider both p-value and effect size in large datasets \citep{10.1111/tri.13535, 10.4300/jgme-d-12-00156.1}. However, the debate (on forums like \emph{StackExchange}) continues over whether huge samples in practice inflate traditional methods' p-values, which is the focus of this work.

Our experiences in differential expression (DE) analysis of single-cell RNA-sequencing (scRNA-seq) data underscore the prevalent concerns regarding inflated p-values. DE analysis, a fundamental aspect of genomic data analysis, aims to identify genes with significantly different expression levels under various conditions \citep{10.1093/bib/bby067}. The scRNA-seq is a high-throughput technique capturing gene expression patterns at the granularity of individual cells. Over the past eight years, scRNA-seq sample sizes have escalated dramatically, growing from a single cell to hundreds of thousands \citep{10.1038/nprot.2017.149}. Most DE methods developed for scRNA-seq data emerged at the inception of this new technology when the number of cells was either small or moderate. Yet, contemporary scRNA-seq data analysis with these DE methods identifies thousands of significant genes (even after multiple testing adjustments), resulting from a large sample size. This has led researchers to rely on subjective log-fold change thresholds (such as 0.5, 1, or 2) to discern differentially expressed genes, presenting two critical problems. First, given the relationship between effect sizes and p-values, the p-values are nearly obsolete for identifying DE genes. Second, it raises the question of whether these numerous small p-values indicate minor effects or if some are misleadingly insignificant due to the large sample sizes involved.

We argue that large samples don't inherently cause issues, but can amplify pre-existing minor problems to critical levels, demanding careful handling. In practice, the ideal conditions posited by statistical theory are rarely encountered, leading to complex issues. Such problems, stemming from unmet model assumptions or data collection flaws, may become more pronounced with larger samples, indicating the necessity for novel strategies to tackle these challenges. To address these challenges, we propose a permutation-based method for deriving p-values unaffected by sample size, even in non-ideal conditions. We utilize a permutation approach since it is flexible to be applied to samples from any distributions \citep{10.2307/2983647,10.2307/2332008}. 

Our simulation study illustrates how sample size affects Type I errors in conventional methods and demonstrates our approach's corrective capabilities. In this study, we deliberately violate model assumptions in two common real-world scenarios. Firstly, we create data with the outcome variable following a distribution different from the model assumption, inspired by the fact that all distribution assumptions in DE analysis over-simplify the intricate nature of real-world data. Secondly, we simulate outcome variable values using two predictors but only include one in the analysis, reflecting situations where not all influencing factors are measurable or considered. This approach mirrors the complexities often encountered in real-world scenarios where comprehensive data capture is challenging.

\section{Notations and Problem Setup}
To investigate how large sample size affects Type I error of statistical analysis with a simulation study, we consider the problem of fitting a Poisson regression to analyze a dataset comprising one outcome variable and one predictor. The objective is to determine the statistical significance of the predictor's association with the outcome variable, primarily through the p-value of the regression coefficient for the predictor. The model is structured as follows:
\begin{align}
\label{e:analysis}
y_i &\sim \mathrm{Poisson}(\lambda_i = e^{\beta_0 + \beta_1 x_{1i}})
\end{align}
Here, $i=1,\ldots, n$ denotes the sample index. The term $y_i$ represents the outcome following a Poisson distribution with parameter $\lambda_i$, and $x_{1i}$ is the predictor for the $i$-th sample.

In situations involving large sample sizes, the concern for statistical power diminishes, shifting our attention primarily to the Type I error rate. To investigate this, we set \(\beta_1=0\) and focus on the proportion of false positives detected in our analysis. Our simulation study considers two specific scenarios, representing two common issues encountered in real-world data analysis.

Firstly, we explore a scenario of distribution misspecification. Here, rather than using a Poisson distribution, the outcomes \(y_i\) are derived from a discretized \(F\) distribution with degrees of freedom \((8,8)\). These values are subsequently rounded to the nearest integer:
\begin{align}
\label{e:generating1}
z_i &\sim F_{8,8} \quad \mathrm{and} \quad y_i = \mathrm{round}(z_i)
\end{align}

Secondly, we consider a scenario involving unobserved predictors. In this case, the outcome \(y_i\) is influenced by an unobserved predictor \(x_{2i}\), modelled as follows:
\begin{align}
\label{e:generating2}
y_i &\sim \mathrm{Poisson}(\lambda_i = e^{\beta_0 + \beta_2 x_{2i}}).
\end{align}

For each scenario, various settings with differing parameter values will be examined. To estimate the Type I error for each setting, we simulate $K= 1000$ datasets and fit the Poisson regression model as per (\ref{e:analysis}). Denote the p-values for $\beta_1$ as $p_j$ for $j=1, \ldots, K$. The Type I error rate for a given setting is then defined as:
\begin{align} \label{e:t1error}
\mathrm{Type\ I\ error\ rate} &= \frac{\sum_{j=1}^K I(p_j<0.05)}{K}
\end{align}
When a method accurately maintains the Type I error rate at the nominal level of 5\%, the decision to reject the null hypothesis in any given simulated dataset follows a Bernoulli distribution with the parameter \(p=0.05\). Consequently, the total count of rejections across \(K=1000\) repeated experiments adheres to a binomial distribution, specifically Binom\((1000, p=0.05)\). This binomial distribution yields a 95\% confidence interval for the estimated Type I error, ranging between (0.0362, 0.0638), which provides a statistical measure of the variability expected around the nominal 5\% Type I error rate. Our simulation study aims to assess the impact of increasing sample sizes on the defined Type I error.

\section{Simulation Study}
\subsection{Impact of Large Sample Size on the Estimated Parameters of a Distribution}
Before tackling the Poisson regression challenge, we first investigate a simpler problem to demonstrate how increased sample sizes can influence the parameter estimation of a Poisson distribution when data collection is flawed. We consider a Poisson distribution where each event count \(y_i\) follows \(y_i \sim \mathrm{Poisson}(\lambda=5)\) for \(i=1, \ldots, n\). However, due to the limited sensitivity of the measuring device, any counts less than 2 are erroneously recorded as 0. This situation sets the stage for understanding the repercussions of discrepancies between the assumed and actual distributions as the number of observations, \(n\), increases.

We anticipate that the measuring device's insensitivity will introduce a bias in estimating \(\lambda\), and we aim to analyze how this bias shifts with enlarging sample sizes. To do this, we simulate datasets of 10 varying sample sizes equally spaced in log-10 scale, i.e. $\mathrm{log}_{10}(n)$ takes 10 values equally spaced between 1 and 6. Each dataset is generated under the flawed measurement condition as described above; from these, we estimate the Poisson parameter \(\hat{\lambda}=\sum_{i=1}^n y_i/n\). We focus on examining the relationship between the sample size \(n\) and the estimation bias \(\hat{\lambda} - \lambda\). To assess the uncertainty in our estimates, we replicate the simulation 1000 times for each sample size, yielding 1000 bias measurements per size.

\begin{figure}
\begin{center}
\includegraphics[width=\textwidth]{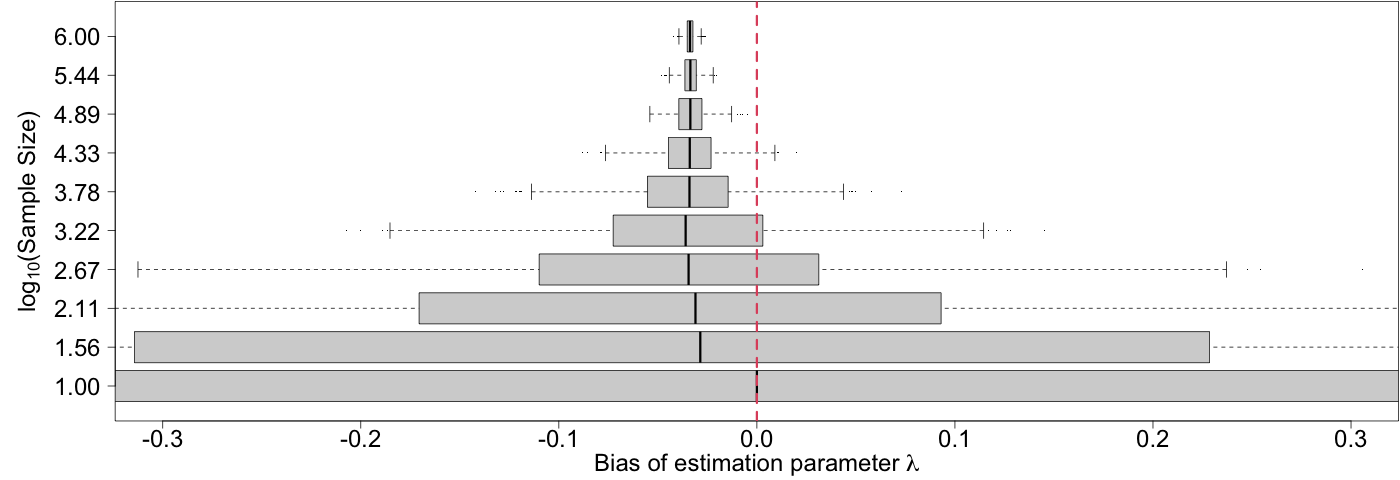}
\end{center}
\caption{Each boxplot displays the distribution of 1000 biases linked to the estimated Poisson rate parameter \(\lambda\) for a specific sample size, indicated on the y-axis. The x-axis denotes the magnitude of these biases, with a dashed red line at 0 indicating the point of unbiased estimation. The boxplots reveal a decrease in estimation uncertainty as the sample size grows, indicated by the shrinking interquartile ranges (IQRs) or the size of boxes. Nevertheless, while precision improves, it also introduces a systematic positive bias that consistently emerges with larger sample sizes. \label{f:poisPara}}
\end{figure}

Each boxplot in Figure~\ref{f:poisPara} displays the distribution of 1000 biases associated with the estimated Poisson rate parameter \(\lambda\) for a specific sample size. The x-axis denotes the magnitude of these biases, with a dashed red line at 0 indicating the point of unbiased estimation. The boxplots reveal a decrease in estimation uncertainty as the sample size grows, indicated by the shrinking interquartile ranges (IQRs) or the size of boxes. However, this increased precision comes with a systematic positive bias that persists across larger sample sizes. In more complicated analyses, this bias in parameter estimation could be consequential, affecting the accuracy of confidence intervals and p-values. For smaller sample sizes, the wider confidence intervals tend to mitigate the effects of bias, reducing its influence on the analysis. However, as sample sizes increase and confidence intervals consequently narrow, even small systematic biases can become critical, leading to substantial errors in inferential decision-making. This emphasizes the importance of accounting for and addressing bias, especially in large-scale data analyses. Next, we will show how growing the sample size affects Poisson regression.

\subsection{Impact of Large Sample Size on Type I Error in Poisson Regression with Assumption Violations}
We now return to the primary focus of this simulation study: Poisson regression. Hypothesis testing of Poisson regression's coefficient is more complex than the basic estimation of a Poisson distribution's parameter, as it involves multiple variables (especially with more than one predictor) and necessitates the determination of a null distribution for the test statistics of the regression coefficients.

Using simulation studies, we explore two distinct scenarios of assumption violations previously outlined in the model framework section.  Here, we fit a Poisson regression model, as defined in Equation~(\ref{e:analysis}), but model assumptions are violated in two different ways. For both scenarios, \(y\) and \(x_1\) are generated as unrelated by setting $\beta_1=0$, making any significant p-values indicative of Type I errors instead of detecting small effects. We aim to examine the influence of increasing sample size on Type I error by analyzing various settings with 60 different sample sizes. Specifically, we let $\mathrm{log}_{10}(n)$ takes 30 values equally spaced between 1 and 2, and another 30 equally spaced between 2 and 5. For each sample size setting, we simulate \(K=1000\) datasets to estimate the corresponding Type I error rate using the formula provided in Equation~(\ref{e:t1error}). More densed sample sizes are investigated in the range $(10^1, 10^2)$ than the range $(10^2, 10^5)$ for two reasons: (1) smaller sample sizes take less computing resources and (2) type I error rate charge faster in the first region in our simulation study.

\begin{figure}
\begin{center}
\includegraphics[width=\textwidth]{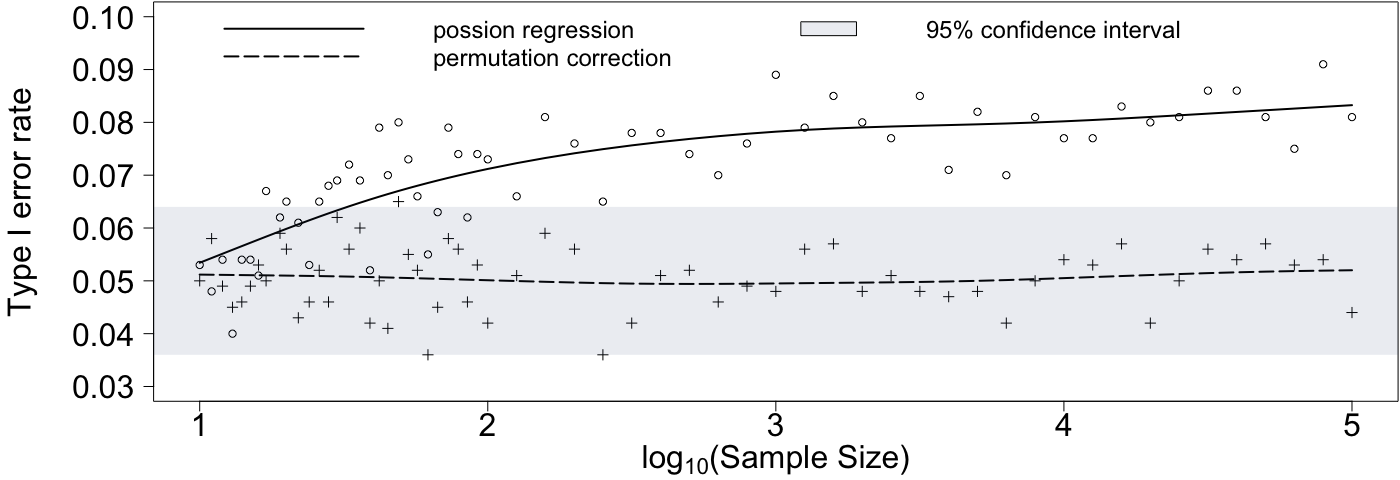}
\end{center}
\caption{The relationship between the sample sizes (x-axis) and the Type I error rates (y-axis), demonstrating how the issue of misspecified outcome distribution is amplified by growing sample sizes. Each circle represents a Type I error rate of Poison regression estimated from \(K=1000\) simulated datasets, and the solid curve represents smoothed pattern of these circles. Each cross represents a Type I error rate of permutation-corrected Poison regression estimated from the same \(K=1000\) simulated datasets, and the dashed curve represents smoothed pattern of these crosses. Shaded area represents the 95\% confidence interval of observed Type I error rate if the method can control it at the nominal level. \label{f:wrongDistn}}
\end{figure}

\textbf{In the first scenario, the actual distribution of the outcomes diverges from the Poisson distribution that the model presumes.} Specifically, the outcomes \(y_i\) are generated from a discretized \(F\) distribution with degrees of freedom \((8,8)\), as detailed in Equation~(\ref{e:generating1}). Predictor data \(x_{1i}\) are drawn from a standard normal distribution \(N(0,1)\). Figure~\ref{f:wrongDistn} illustrates the relationship between the sample sizes (x-axis) and the Type I error rates (y-axis), demonstrating how the issue of misspecified outcome distribution is amplified by growing sample sizes. Each circle represents a Type I error rate of Poison regression estimated from \(K=1000\) simulated datasets, and the solid curve represents smoothed pattern of these circles. Shaded area represents the 95\% confidence interval of observed Type I error rate if the method can control it at the nominal level. The estimated Type I error rates of Poison regression remain within the 95\% confidence bounds when the sample size is small. As sample size increases Type I error rates start to fluctuate in and out of the confidence bounds, reflecting inconsistent control over false positives. When the sample size keep growing, the estimated Type I error predominantly lies outside these bounds, indicating that larger samples exacerbate Type I error inflation when models use wrong distribution assumptions. The Type I error rates become stable at certain values when sample size is large enough, which converges to the theoratical value depends on how the model assumption is voilated. 

Both the discretized \(F\) distribution and the Poisson distribution are integer-valued and skew towards the right, possessing heavier tails on that side. Our simulation study indicates that when the sample size is small, a misspecified distribution poses a minor problem, and approximating a Poisson distribution with a discretized F distribution is generally acceptable. However, as the sample size increases, the ability to differentiate between the two distributions enhances, thereby aggravating the issue of distribution misspecification. This escalation becomes critical, leading to incorrect decisions, notably inflated Type I errors.

\begin{figure}
\begin{center}
\includegraphics[width=\textwidth]{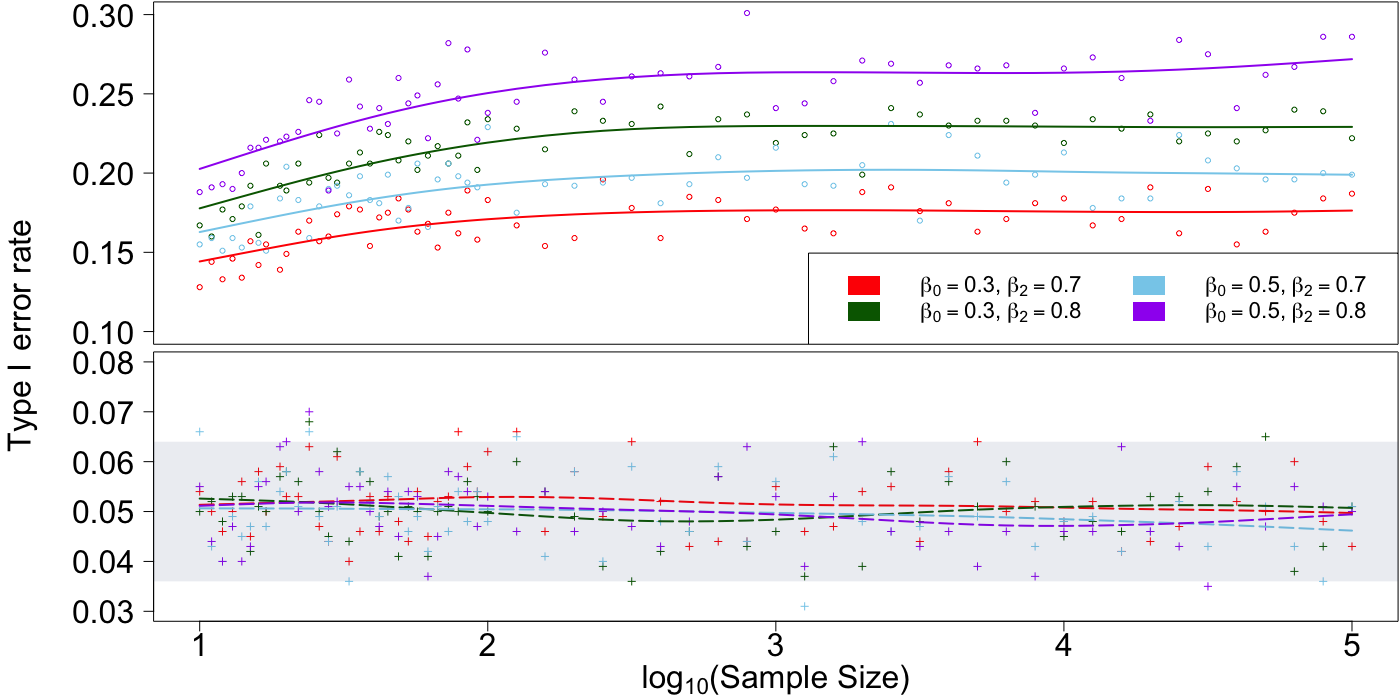}
\end{center}
\caption{The relationship between the sample sizes (x-axis) and the Type I error rates (y-axis), demostrating how the impact of missing a predictor can be amplified by growing sample sizes. Each circle represents a Type I error rate of Poison regression estimated from \(K=1000\) simulated datasets, and the solid curve represents smoothed pattern of these circles. Each cross represents a Type I error rate of permutation-corrected Poison regression estimated from the same \(K=1000\) simulated datasets, and the dashed curve represents smoothed pattern of these crosses. Shaded area represents the 95\% confidence interval of observed Type I error rate if the method can control it at the nominal level. Four coefficient settings are represented by different colors, as annotated in the figure legend. \label{f:missVar}}
\end{figure}

\textbf{In the second scenario, outcomes influenced by an unobserved predictor.} Specifically, \(y_i\) are generated using \(x_{2}\) according to the true model specified in Equation~(\ref{e:generating2}). Consequently, \(y\) and \(x_1\) share no relationship, and the actual predictor \(x_2\) is not included in the data analysis model described in Equation~(\ref{e:analysis}). In this setup, we simulate both the noise predictor \(x_{1i}\) and the genuine predictor \(x_{2i}\) from a standard normal distribution \(N(0,1)\). We explore four different settings of true regression coefficient values in model~(\ref{e:generating2}), particularly selecting combinations of \(\beta_0=0.3, 0.5\) controlling the magnitude of baseline Poison rates and \(\beta_2=0.7, 0.8\) controlling the strength of unobserved signal. Figure~\ref{f:missVar} illustrates the relationship between the sample sizes (x-axis) and the Type I error rates (y-axis). Same as in the first scenario, each circle represents a Type I error rate of Poison regression estimated from \(K=1000\) simulated datasets, and the solid curve represents smoothed pattern of these circles. Shaded area represents the 95\% confidence interval of observed Type I error rate if the method can control it at the nominal level. Four coefficient settings are represented by different colors, as annotated in the figure legend. Notably, we observed that the Type I errors in all settings exceeded the 95\% confidence interval's upper bound 0.0638, indicating a severe inflation. In this scenario, the distribution of outcomes \(y_i\) is influenced by two factors: the actual counts that follow a Poisson distribution and Gaussian noise attributable to the unobserved \(x_2\), which is not accounted for in the model and consequently contributes to the residual variation. This omission of a variance source leads to overconfident inferences, which explains the exceedingly high Type I errors, even with as few as 10 samples. The simulation study reveals that the impact of this omission intensifies as the sample size increases up to $n=100$. Beyond this point, the impact begins to plateau, ultimately aligning with the theoretical implications of disregarding a normal noise component in the analysis. Upon comparing results across four parameter settings, we observed that higher values of \(\beta_2\)—indicative of stronger missing signals—correspond to more severe inflation of Type I error rates. This is consistent with the notion that the greater the unobserved information, the more compromised the analysis results are. The variance in the data originates from two components: the inherent randomness in the Poisson distribution, governed by \(\beta_0\), and the variation in the Normal distribution of unobserved predictor, dictated by \(\beta_2\). Poisson regression uses Poison distribution to encapsulate uncertainty and fails to distinguish between these two noise sources. Consequently, we also observed that increased \(\beta_0\) values, which amplify the variance attributed to the Poisson component, further exacerbate the inflation of Type I error rates when there is an unobserved predictor.

Comparing results among four parameter settings, we found larger $\beta_2$ values (i.e. missing stronger signals) lead to more severe inflation of Type I error rates, which agrees with our intuition that the more unobserved information, the worse analysis results. The variance in data comes from two sources: the randomness in Poison distribution controlled by parameter $\beta_0$ and in Normal distribution controlled by $\beta_2$. When the model only uses Poisson distribution to model the uncertainty, noises from two sources cannot be isolated; hence, we also found that larger $\beta_0$ values (increasing Poison variance) lead to more severe inflation of Type I error rates.

\section{Using Permutation-Based Methods to Address Issues Exacerbated by Large Sample Sizes}
Our simulation studies demonstrate that large sample sizes can inflate the Type I error rate when the model assumptions of Poisson regression are violated, either due to outcome distribution misspecification or the presence of unobserved predictors. When the sample size is large enough, the damage of assumption violations converges to its theoretical level. These issues are rooted in an incorrect theoretical null distribution. To counteract these problems, we propose a permutation-based approach to construct a robust null distribution for the test statistic, specifically \(\hat{\beta}_1\), the estimate from Poisson regression. This method aims to ascertain the distribution of \(\hat{\beta}_1\) under the null hypothesis of no relationship between \(x_1\) and \(y\).

Initially, we fit a Poisson regression model to the original data to obtain the observed coefficient, \(\hat{\beta}_1^{orig}\). We then create numerous permuted versions of the original dataset (e.g., \(N=1000\)) by randomly shuffling the values of \(\mathbf{x}_1\) to break any actual connection between \(x_1\) and \(y\). For each permutation, we re-estimate \(\hat{\beta}_1\) after fitting the Poisson regression model to the permuted data, generating a new coefficient, \(\hat{\beta}_{1,perm}\). These coefficients from all permutations form an empirical null distribution for \(\hat{\beta}_1\) under the hypothesis of no association.

Upon completing all permutations, the p-value is computed as the proportion of permuted datasets where the absolute value of the permuted coefficient equals or exceeds that of the original coefficient. This p-value indicates the probability of observing the given association between \(x_1\) and \(y\) if, in fact, no such association exists.

This permutation-based method offers robustness against model misspecification and unobserved predictors. Issues affecting the original dataset equally impact the permuted datasets used to construct the null distribution. Moreover, as the sample size increases, it does so for both the original and permuted data, thereby neutralizing the effects of assumption violations and sample size influences. The entire process is succinctly encapsulated in Algorithm~\ref{alg:permTest}, offering a systematic approach to addressing the challenges of large sample sizes in statistical analysis.

\begin{algorithm}
\caption{Permutation-based p-value for Poisson Regression}\label{alg:permTest}
\begin{algorithmic}[1]

\Require \(\mathbf{y}=(y_1, \ldots, y_n) \), \(\mathbf{x}_{1}= (x_{1i},\ldots, x_{1n})\), \(N\) (default \(N = 1000\))
\State Fit the Poisson regression model \(y \sim \mathrm{Poisson}(\lambda = e^{\beta_0 + \beta_1 x_1})\) to the data \(\mathbf{y}\) and \(\mathbf{x}_{1}\).
\State Record the estimate \(\hat{\beta}_1^{orig}\) from the original data.
\State Initialize a counter \(count = 0\) for the number of times permuted \(\beta_1\) exceeds the original.
\For{\(j = 1\) to \(N\)} 
    \State Permute \(x_{1i}\) to get \(x_{1,perm}^{(j)}\).
    \State Fit the model to the permuted data \(y \sim \mathrm{Poisson}(\lambda = e^{\beta_0 + \beta_{1,perm} x_{1,perm}^{(j)}})\).
    \State Compute the estimate \(\hat{\beta}_{1,perm}^{(j)}\) from the permuted data.
    \If{\(|\hat{\beta}_{1,perm}^{(j)}| \geq |\hat{\beta}_1^{orig}|\)}
    \(count \leftarrow count +1 \).
    \EndIf
\EndFor
\State Compute the p-value as \(p = count / N\).
\State \textbf{return} p-value \(p\).

\end{algorithmic}
\end{algorithm}

We implemented the permutation-based algorithm on data generated in our simulation study. The crosses and dashed curves in Figures~\ref{f:wrongDistn} and \ref{f:missVar} show estimated Type I errors when using our permutation-based approach to analyze the same datasets generated in the Simulation Section. We discovered that our method effectively corrects the inflated Type I errors of the original Poisson regression to mostly within the shaded area representing 95\% confidence interval the nominal level of \(0.05\). Furthermore, the dashed trend curves of all four settings fluctuate around 0.05 with no discernible trend between the sample sizes and the Type I errors rates obtained through permutation correction. This indicates the robustness of this method in maintaining error rates in big data.

\section{Conclusion and Discussion}
In our simulation studies, we opted to permute \(x_1\) rather than \(y\), a decision that is not critical when the analysis involves only two variables. However, when multiple predictors are incorporated into the regression analysis, permuting \(x_1\) becomes essential. This is because our primary objective is to scrutinize the relationship between \(x_1\) and \(y\), while accounting for the effects of other predictors. Permuting \(y\) in such contexts would disrupt its relationship not only with \(x_1\) but also with the other covariates. Consequently, the null distribution derived from permuting \(y\) would no longer accurately reflect the true relationships between \(y\) and the other covariates when it's assumed to be unrelated to \(x_1\). Therefore, the specific choice of permutation is crucial in maintaining the integrity and interpretability of the analysis, especially in more complex regression models with multiple predictors.

In our simulation studies, we observed that the detrimental effects of model assumption violations become more pronounced with increasing sample sizes, yet the extent of these damages diminishes as the sample size continues to grow. This observation parallels the well-known phenomenon where the benefits of enhanced statistical power from increasing sample size diminish at larger scales. Both phenomena stem from the same underlying principle: as the sample size approaches infinity, the information extracted from the data asymptotically approaches the theoretical truth of the data's characteristics. Consequently, both the benefits and the detriments of analysis are bound to converge towards their theoretical limits. Therefore, when the current value is near this limit, further enlarging the sample size yields minimal changes. This principle underscores the diminishing returns of increasing sample size, both in terms of benefits and in terms of mitigating the effects of assumption violations.

The Poisson regression model used in our simulation study exemplifies how certain statistical analyses necessitate specialized methods to uphold nominal Type I error rates when dealing with large-sample data. Nevertheless, it's important to recognize that not all analytical methods require such permutation-based corrections. For instance, the Central Limit Theorem (CLT) posits that the sample mean converges to a normal distribution as the sample size increases, assuming certain conditions are met \citep{10.1214/ss/1177013818}. Consequently, if the null distribution of a test is derived based on the assumption that the sample mean follows a normal distribution, such methods do not require correction in large-sample data analysis. This highlights the nuanced nature of statistical methodologies, where the need for adjustments like permutation-based remedies is contingent upon the specific assumptions and characteristics of the test in question.

In conclusion, the correct p-value from some statistical methods relies on correctly specifying the model and including all relevant variables. Sample size increase may exacerbate the problem caused by incorrect model assumptions or flaws in data collection. So, when analyzing large sample data, special treatment may be necessary to obtain the correct p-values. We proposed a permutation-based p-value that is robust to many types of model misspecification and unobserved confounders, and its p-values are unaffected by sample sizes. This makes it a powerful tool, especially in complex real-world scenarios where assumptions about the data or model may not fully hold and the sample size is large.

%% if your bibliography is in bibtex format, uncomment commands:
\bibliographystyle{chicago}
\bibliography{jcgs-submission}       % Bibliography file (usually '*.bib')

\begin{thebibliography}{}

\bibitem[\protect\citeauthoryear{Armstrong}{Armstrong}{2019}]{10.1111/opo.12618}
Armstrong, R.~A. (2019).
\newblock {Is there a large sample size problem?}
\newblock {\em Ophthalmic and Physiological Optics\/}~{\em 39\/}(3), 129--130.

\bibitem[\protect\citeauthoryear{Cam}{Cam}{1986}]{10.1214/ss/1177013818}
Cam, L.~L. (1986).
\newblock {The Central Limit Theorem Around 1935}.
\newblock {\em Statistical Science\/}~{\em 1\/}(1), 78--91.

\bibitem[\protect\citeauthoryear{Dunkler, Haller, Oberbauer, and Heinze}{Dunkler et~al.}{2020}]{10.1111/tri.13535}
Dunkler, D., M.~Haller, R.~Oberbauer, and G.~Heinze (2020).
\newblock {To test or to estimate? P‐values versus effect sizes}.
\newblock {\em Transplant International\/}~{\em 33\/}(1), 50--55.

\bibitem[\protect\citeauthoryear{Faber and Fonseca}{Faber and Fonseca}{2014}]{10.1590/2176-9451.19.4.027-029.ebo}
Faber, J. and L.~M. Fonseca (2014).
\newblock {How sample size influences research outcomes}.
\newblock {\em Dental Press Journal of Orthodontics\/}~{\em 19\/}(4), 27--29.

\bibitem[\protect\citeauthoryear{McDermaid, Monier, Zhao, Liu, and Ma}{McDermaid et~al.}{2018}]{10.1093/bib/bby067}
McDermaid, A., B.~Monier, J.~Zhao, B.~Liu, and Q.~Ma (2018).
\newblock {Interpretation of differential gene expression results of RNA-seq data: review and integration}.
\newblock {\em Briefings in Bioinformatics\/}~{\em 20\/}(6), 2044--2054.

\bibitem[\protect\citeauthoryear{Pitman}{Pitman}{1937}]{10.2307/2983647}
Pitman, E. J.~G. (1937).
\newblock {Significance Tests Which May be Applied to Samples from Any Populations. II. The Correlation Coefficient Test}.
\newblock {\em Supplement to the Journal of the Royal Statistical Society\/}~{\em 4\/}(2), 225--232.

\bibitem[\protect\citeauthoryear{Pitman}{Pitman}{1938}]{10.2307/2332008}
Pitman, E. J.~G. (1938).
\newblock {Significance Tests which May be Applied to Samples from any Populations: III. The Analysis of Variance Test}.
\newblock {\em Biometrika\/}~{\em 29\/}(3/4), 322.

\bibitem[\protect\citeauthoryear{Sullivan and Feinn}{Sullivan and Feinn}{2012}]{10.4300/jgme-d-12-00156.1}
Sullivan, G.~M. and R.~Feinn (2012).
\newblock {Using Effect Size—or Why the P Value Is Not Enough}.
\newblock {\em Journal of Graduate Medical Education\/}~{\em 4\/}(3), 279--282.

\bibitem[\protect\citeauthoryear{Svensson, Vento-Tormo, and Teichmann}{Svensson et~al.}{2018}]{10.1038/nprot.2017.149}
Svensson, V., R.~Vento-Tormo, and S.~A. Teichmann (2018).
\newblock {Exponential scaling of single-cell RNA-seq in the past decade}.
\newblock {\em Nature Protocols\/}~{\em 13\/}(4), 599--604.

\end{thebibliography}

%% or include bibliography directly:
% \begin{thebibliography}{}
% \bibitem{b1}
% \end{thebibliography}

\end{document}